\documentclass{appolb}
\usepackage{graphicx}

\begin{document}
\title{Final combined deep inelastic scattering cross sections\\ at HERA%
\thanks{Presented at Excited QCD 2016 Workshop.}%
}
\author{Matthew Wing \\(on behalf of the H1 and ZEUS Collaborations)
\address{Department of Physics and Astronomy, University College London}
}
\maketitle
\begin{abstract}
The combination is presented of all inclusive deep inelastic scattering cross sections previously published by the H1 and 
ZEUS collaborations at HERA for neutral and charged current $ep$ scattering for zero beam polarisation. The data were 
taken at proton beam energies of 920, 820, 575 and 460\,GeV and an electron beam energy of 27.5\,GeV.  The data 
correspond to an integrated luminosity of about 1\,fb$^{-1}$ and span six orders of magnitude in negative 
four-momentum-transfer squared, $Q^2$, and Bjorken $x$. The correlations of the systematic uncertainties were evaluated 
and taken into account for the combination. The combined cross sections were input to QCD analyses at leading order, 
next-to-leading order and at next-to-next-to-leading order, providing a new set of parton distribution functions, called 
HERAPDF2.0.  Additionally, the inclusion of 
jet-production cross sections made a simultaneous and precise determination of 
parton distributions and the strong coupling constant possible.  Brief highlights of the results are presented.
\end{abstract}
  
\section{Introduction}

Measurements of deep inelastic lepton--nucleon scattering have long been a way of probing the dynamics of the strong force, 
yielding a detailed picture of the structure of the proton and precise extractions of the strong coupling constant.  This is 
important in trying to understand the structure of matter at its most basic level, but also has direct application to other projects 
where nucleons are or will be used such as the LHC.  Data from HERA provides the backbone of our understanding of 
the structure of the proton with numerous measurements performed to high precision and covering a wide kinematic range.  
All data from the H1 and ZEUS Collaborations on measurements of inclusive deep inelastic scattering have recently been 
combined~\cite{epj:c75:580}.  These data will provide the basis for understanding the structure of the proton for many years to come.  
The data provide many beautiful demonstrations of fundamental physical phenomena such as scaling violations of the cross 
section and electroweak unification.  The high precision also allows their sole use in a QCD fit to determine the parton distribution 
functions in the proton (called HERAPDF2.0).

\section{Data combination}

The data, taken over the 15-year lifetime of the HERA accelerator, correspond to a total luminosity of about 1\,fb$^{-1}$ of deep 
inelastic electron--proton and positron--proton scattering.  All data used were taken with an electron beam energy of 27.5\,GeV.  
Roughly equal amounts of data for electron--proton and positron--proton scattering were recorded.  
The bulk of the data were taken with a proton beam energy of 920\,GeV, but samples with proton beam energies of 820, 575 and 
460\,GeV were also collected.  The data were combined separately for the $e^+p$ 
and $e^-p$ data and the different centre-of-mass energies.  Overall, 41 separate data sets were used in the combination, spanning 
the ranges $0.045 < Q^2 < 50\,000$\,GeV$^2$ and $6 \times 10^{-7} < x_{\rm Bj} < 0.65$, i.e.\ six orders of magnitude in each 
variable.  The initial measurements consisted of in total 2937 published cross sections which were combined to 1307 final 
combined cross-section measurements.  

\begin{figure}[htb]
\centerline{%
\includegraphics[width=10.5cm]{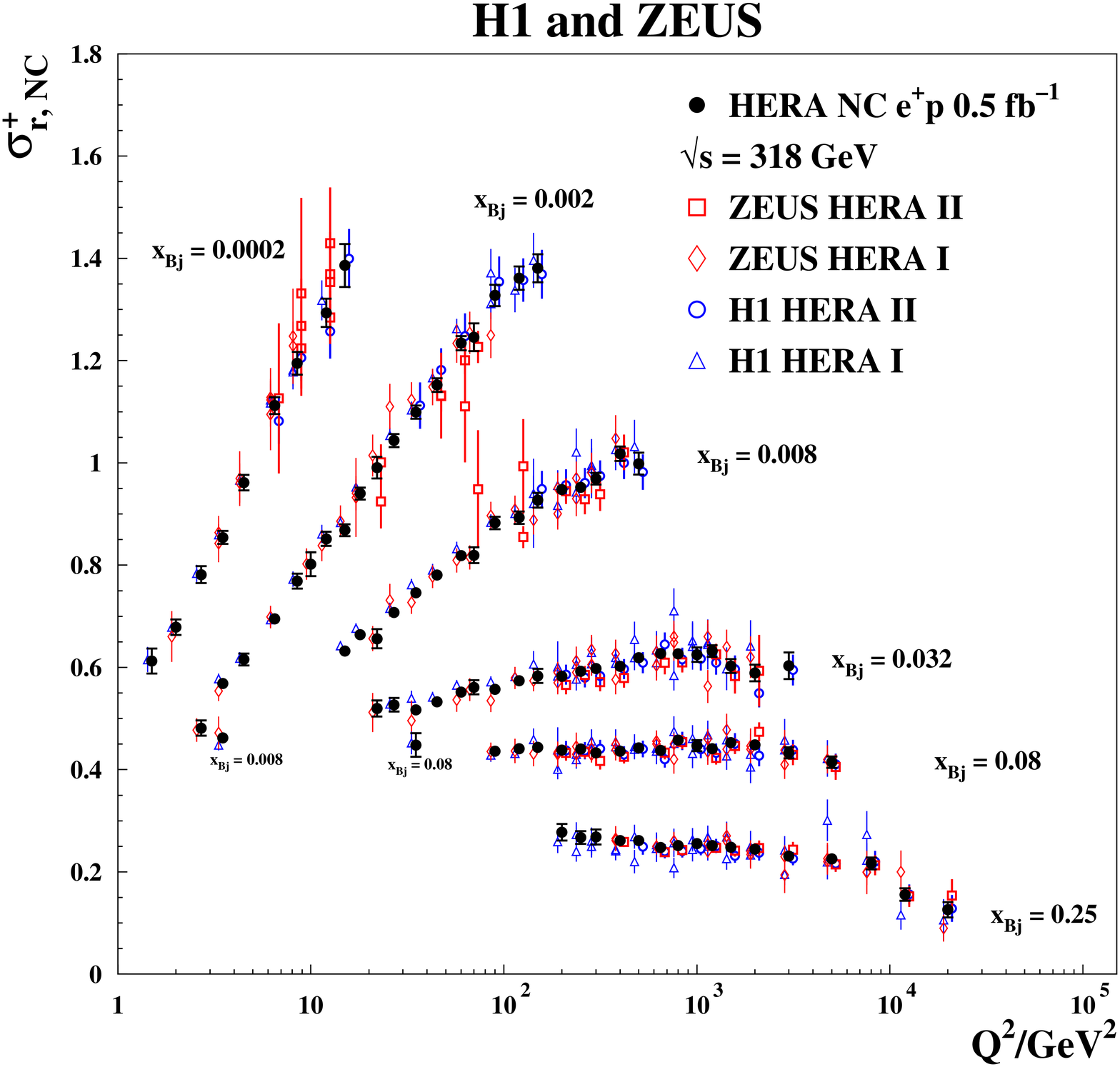}}
\caption{Neutral current reduced cross section, $\sigma_{\rm r,\,NC}^+$, versus $Q^2$ for H1 and ZEUS
  data sets (open points) and after the HERA combination (solid points).  A selection of
  data for fixed $x_{\rm Bj}$ is shown.}
\label{fig:data-comb}
\end{figure}

The data combination procedure involved a careful treatment of the various uncertainties between all the sets of data.  In 
particular the correlations of the various sources were assessed and those uncertainties deemed to be point-to-point correlated 
were accounted for as such in the averaging of the data based on a $\chi^2$ minimisation method.  The resulting $\chi^2$ is 1687 
for 1620 degrees of freedom, demonstrating excellent compatibility of the multitude of data sets.  The power of the data combination 
can be seen in Fig.~\ref{fig:data-comb} which displays a selection of the data in bins of the photon virtuality, $Q^2$, and for fixed 
values of Bjorken $x_{\rm Bj}$.  The individual data sets from several different publications are shown here separately. A combined 
data point can be the combination of up to 8 individual measurements and the improvement in precision is striking.  An indication 
of the precision of the combined data is that the total uncertainties are close to 1\% for the bulk region of $3 < Q^2 < 500$\,GeV$^2$.

\section{Physics highlights}

The data in Fig.~\ref{fig:data-comb} also demonstrate beautifully the effect of scaling violations in deep inelastic scattering.  
At $x_{\rm Bj} \sim 0.1$, the cross section is flat with $Q^2$, whereas the cross section falls with increasing $Q^2$ at high $x_{\rm Bj}$ 
and rises rapidly with increasing $Q^2$ at low $x_{\rm Bj}$.  This rapid rise is indicative of an ever-increasing gluon density 
being probed, shown here for a subset of the data sample.

\begin{figure}[htb]
\centerline{%
\includegraphics[width=10.5cm]{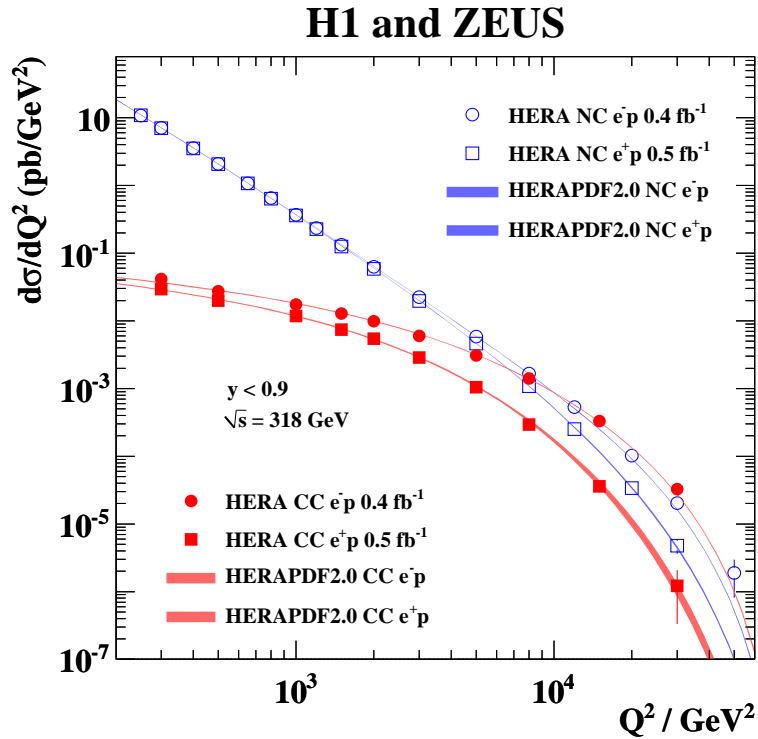}}
\caption{Neutral current (open points) and charged current (solid points) differential cross sections
versus $Q^2$ compared to Standard Model predictions (solid curves).}
\label{fig:ew-uni}
\end{figure}

The differential cross sections at high $Q^2$ are shown in Fig.~\ref{fig:ew-uni} for neutral and charged current events for both 
$e^+p$ and $e^-p$ interactions.  At $Q^2 \sim 200$\,GeV$^2$, the neutral current cross section is significantly higher than the charged 
current cross section due to the dominance of photon exchange.  At $Q^2 > 10\,000$\,GeV$^2$, the cross sections become similar, 
due to the exchange of massive vector bosons, indicative of the unification of the electromagnetic and weak forces.  The difference in 
the neutral current cross sections for $e^+p$ and $e^-p$ data is due to $\gamma-Z$ interference effects; this allowed an extraction of 
the $xF_3$ structure function which is related to the density of valence quarks (not shown).  The helicity structure of the $W^\pm$ 
exchange and the different quarks being probed also leads to the observed difference in $e^+p$ and $e^-p$ data in charged current 
interactions.

The data demonstrating these fundamental physical properties are well described by the Standard Model, encompassed on the 
QCD fit, HERAPDF2.0, described in the next section.

\section{QCD analysis, HERAPDF2.0}

\begin{figure}[htb]
\centerline{%
\includegraphics[width=10.5cm]{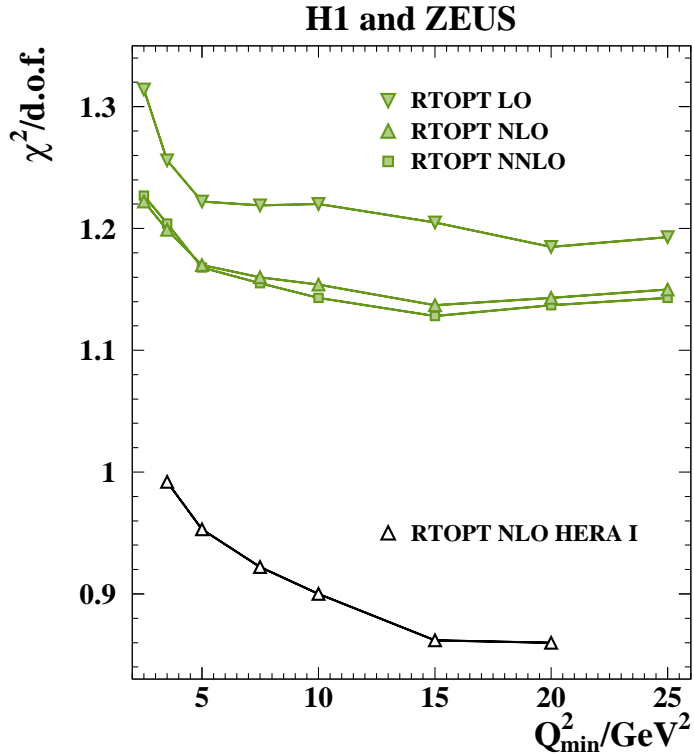}}
\caption{$\chi^2$/d.o.f. of QCD fit to HERA data for differing minimum values of $Q^2$.  Results are shown at different 
orders for the calculation and also for the fit to HERA I data only.  The RTOPT~\cite{rtopt} label signifies the scheme used to treat 
heavy quarks.}
\label{fig:fit_vs_q2}
\end{figure}

The HERA data was then used as the sole input to a QCD analysis using the DGLAP equations at leading (LO), next-to-leading (NLO) 
and next-to-next-to-leading order (NNLO).  The data were restricted to $3.5<Q^2<50\,000$\,GeV$^2$, with the minimum $Q^2$ value 
varied and the effects studied as shown in Fig.~\ref{fig:fit_vs_q2}.  As can be seen in the figure, the overall $\chi^2$/d.o.f. for the full 
$Q^2$ range is about 1.2 (at both NLO and NNLO), but becomes lower until about 10\,GeV$^2$ where is levels out at a value of about 
1.14.  The same trend with $Q^2$, but just with different absolute values was present in the HERA I data (1992--2000)~\cite{herapdf1}.  
This trend towards lower $Q^2$ indicates that something more is needed beyond DGLAP evolution and a full and consistent description 
of these data will lead to a deeper understanding of QCD.  However, it should be noted that predictions for LHC processes do not differ 
between the fits with $Q^2_{\rm min}$ of 3.5\,GeV$^2$ or 10\,GeV$^2$.

\begin{figure}[htb]
\centerline{%
\includegraphics[width=10.5cm]{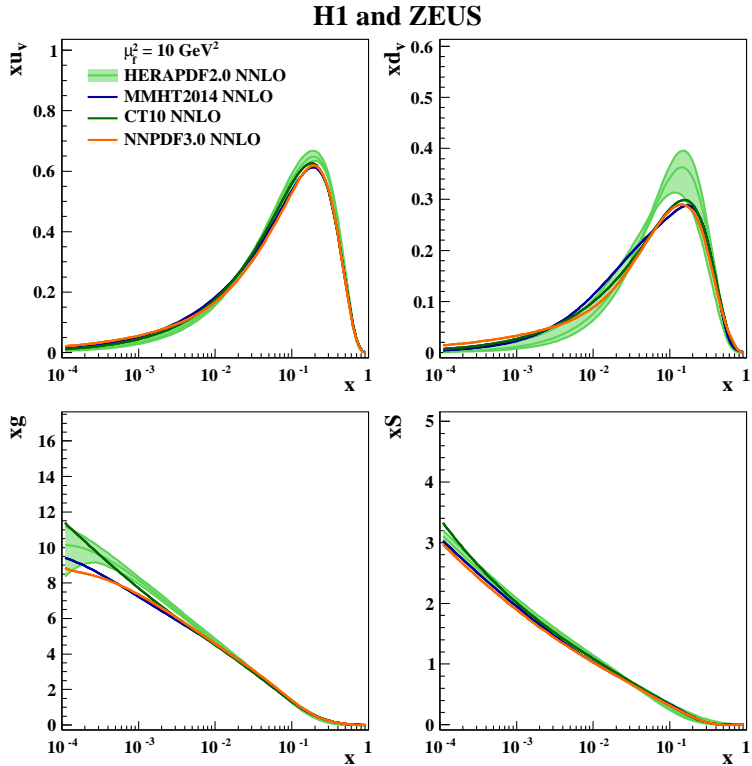}}
\caption{Comparison of HERAPDF2.0 at NNLO with QCD fits from other groups, showing the $u$ and $d$ valence, gluon and sea 
parton density functions.}
\label{fig:pdfs}
\end{figure}

A comparison of HERAPDF2.0 is shown with predictions from other groups~\cite{mmht,ct10,nnpdf} in Fig.~\ref{fig:pdfs}.  Overall the 
expectations from different groups are compatible.

Jet and charm data were also included in an NLO QCD fit allowing an extraction of the strong coupling constant.  The QCD fit described 
the data well and yielded a value of the strong coupling constant consistent with the world average and with an experimental precision 
of better than 1\%.  The theoretical uncertainty due to varying the renormalisation and factorisation scales was significantly larger, 
about 3\%, and could hopefully be improved with a NNLO jet calculation and so allowing a NNLO fit of the data.

\subsection*{Acknowledgements}

Support from DESY and the Alexander von Humboldt Stiftung are gratefully acknowledged.

\end{document}